\documentclass[12pt]{iopart}
\usepackage{latexsym}
\usepackage{dcolumn}
\usepackage{bm}
\input{epsf}
\newcommand{\be}{\begin{equation}}
\newcommand{\ee}{\end{equation}}
\newcommand{\ba}{\begin{eqnarray}}
\newcommand{\ea}{\end{eqnarray}}
\def\la{\lower0.6ex\vbox{\hbox{$ \buildrel{\textstyle <}\over{\sim}\ $}}}
\def\ga{\lower0.6ex\vbox{\hbox{$ \buildrel{\textstyle >}\over{\sim}\ $}}}

\def\mnras{MNRAS}
\def\apj{ApJ}
\def\physrep{Physics Report}
\def\prd{PRD}

\begin{document}
\title[The ergodicity bias]{The ergodicity bias in the observed galaxy distribution}
\author{Jun Pan$^1$,Pengjie Zhang$^2$}
\address{$^1$The Purple Mountain Observatory, 2 West Beijing Road,
  Nanjing 210008, China} 
\address{$^2$Key Laboratory for Research in Galaxies and Cosmology, Shanghai
  Astronomical Observatory, 80 Nandan Road, Shanghai, 200030, China
}

\ead{jpan@pmo.ac.cn}

\begin{abstract}
The spatial distribution of galaxies we 
observed is subject to the given condition that
we, human beings are sitting right in a galaxy -- the Milky Way. Thus the
ergodicity assumption is questionable in interpretation of the observed
galaxy distribution. The resultant difference between observed 
statistics (volume average) and the true cosmic value (ensemble average) is
termed as the ergodicity  bias. We perform explicit numerical investigation of
the effect for a set of galaxy survey depths and near-end distance cuts. It is 
found that the ergodicity bias in observed two- and three-point correlation
functions in most cases is insignificant for modern analysis of samples from
galaxy surveys and thus close a loophole in precision cosmology. However, it
may become non-negligible in certain circumstances, such as those
applications involving three-point correlation function at large scales of
local galaxy samples. Thus one is reminded to take extra care in galaxy
sample construction and interpretation of 
the statistics of the sample, especially when the characteristic redshift is low.
\end{abstract}


\section{Introduction}
One key support to the Cosmological Principle is the observed
near-isotropy of the cosmic microwave background radiation and the
angular distribution of galaxies. But isotropy alone does not prove
homogeneity, the crucial link from isotropy to
homogeneity is the Copernican Principle, which asserts that we
are not privileged observer sitting in a special place in the Universe.

Then there is the ergodicity assumption which states that by
averaging over sufficiently large volume
the measured statistics (volume average) is equivalent to the
statistics on ensemble average.
It is with the Cosmological Principle and the ergodicity assumption that we
believe for any galaxy survey, as long as its effective volume is 
sufficiently large so that the cosmic variance can be ignored, the resulted 
sample is a fair representation of the 
Universe \cite{Peebles1980,ColesLucchin2002book,Weinberg2008book}.

It is true that there are no proper reasons to resurge the specialty of
human beings in the modern cosmology, although there are works
claiming we are in the center of a giant local void (e.g. \cite{BolejkoWyithe2009}).
Nevertheless, strictly speaking, the validity of the ergodicity still 
requires averaging
over the observer positions to avoid possible selection
bias. Unfortunately, in reality, we are only able to observe the galaxy 
distribution from the Milky Way. We are not statistically different to 
those observers in other galaxies than the Milky way, but we are different to
those observers not residing in any galaxy. The distribution of
galaxies we observed shall be interpreted as the distribution of neighbors to
us. This point is a mathematical one 
rather than philosophy. Namely, we have to evoke the {\em conditional}
statistics given then already existence of the Milky Way in which we live to 
interpret the observed galaxy distribution,
instead of the unconditional ones in compliance with the Copernican principle and 
the ergodicity assumption.

For this reason, we call the difference between the volume averaged
galaxy distribution observed by us and the ensemble average as 
the {\em ergodicity bias}.  This is a previously unknown loophole in precision
cosmology and galaxy statistics. The statistical tools to deal with it
turns out to be the conditional statistics, which are actually all there in
the classical textbook of  \cite{Peebles1980}. 
We will see in the following sections that the change to the way 
of thinking brings interesting conclusions about the
measured galaxy number density, two-point correlation function (2PCF) and the
three-point correlation function (3PCF). 

The idea is not completely new, concern about the fairness of sample
is repeatedly expressed in the book of \cite{Peebles1980}, in which there is the
clear recognition that the accidental perfect galaxy number counts 
Hubble \cite{Hubble1926} achieved is 
partly resulted from {\em ``substantial excess of bright galaxies due to the local 
concentration in and around the Virgo cluster''} (p. 5 of \cite{Peebles1980}).
But to our knowledge, the paper presented here is the first to explicitly address and
numerically evaluate the ergodicity bias. And we do find that the ergodicity bias is
negligible in most cases and thus close a loophole in modern cosmology and
galaxy statistics. However, in some cases especially when the characteristic
redshift is low, one may need to take extra care of this ergodicity bias.

\section{Distribution of galaxies as we observed}

\subsection{Number density}
To see how it comes, first let us check the spatial number density of 
galaxies. Let $n_g({\bf r})$ denotes the local number density of galaxies
at position ${\bf r}$, then there are two averages: the ensemble average
$\langle n_g({\bf r})\rangle_e$ and the spatial average $\langle n_g({\bf r})\rangle_{\mathcal R}$
over sample space ${\mathcal R}$.

By the Cosmological Principle the ensemble average $\langle n_g({\bf r}) \rangle_e=n_0$
is a constant everywhere, while the spatial average 
\begin{equation}
\hat{n}_0=\langle n_g({\bf r})\rangle_{\mathcal R}=
\frac{\int_{\mathcal R} n_g({\bf r}) d{\bf r}}{\int_{\mathcal R} d{\bf r}}
\end{equation}
is not, may depend on the position and the shape of the sample. The ergodicity
assumption then just makes the two equal if the sample space ${\mathcal R}$ is
large enough to suppress cosmic variance, no matter the big volume is achieved by 
depth increment or sky coverage enlargement. 

But a fact is that as we are already in a galaxy, the isotropic radial number density of objects
at distance $r$ to us is a conditional number density and
is expected to be
\begin{equation}
\bar{n}_g(r)=n_0 \left[ 1+ \xi_{Gg}(r)\right] \ ,
\label{eq:dens}
\end{equation}
where $\xi_{Gg}$ is the two-point cross correlation function between
the Galaxy and sample galaxies. Then
the measured mean number density for the sample defined by distance
limit $[r_{\rm min}, r_{\rm max}]$ and sky coverage of $4\pi f$ steradians
is 
\begin{equation}
\hat{n}_0=\langle n_g({\bf r})\rangle_{\mathcal R}=n_0
\frac{4\pi f\int_{r_{\rm min}}^{r_{\rm max}} \left[1+\xi_{Gg}(r)\right] r^2{\rm d}r}
{4\pi f(r_{\rm max}^3 - r_{\rm min}^3 )/3}\ .
\end{equation}
It is very clear that this introduces a systematic bias ascribed to
the long range correlation between the Galaxy with other galaxies,
simply improving the sky coverage can not alleviate the bias.

The integration $\int_{r_{\rm min}}^{r_{\rm max}}\xi_{Gg} r^2 dr$
in generally is not zero except for $r_{\rm min}=0, r_{\rm max}=\infty$ or
well designed pair of distance cuts to force a zero provided that the $\xi_{Gg}$ is 
known already at any desired distance in advance. However it is impossible to
push $r_{max}$ to infinity or always have the luck to
meet with the right pair of distance cuts. The point is that no matter how
deep or large the sample could be, there is the general non-zero systematics of 
$\hat{n}_0-n_0$ regardless how small it is, our spatially averaged mean number density 
does not equal to the ensemble average, though asymptotically approaches, i.e. there is the
ergodicity bias.

As stated in Eq.~\ref{eq:dens} the modulation to the local galaxy number 
density depends on $\xi_{Gg}$, which
calls for caution in taking local galaxy samples for distance-number counts 
related statistics, e.g. the luminosity function:
redshift gradient resulted from $\bar{n}_g$ is in fact incorporated 
into the evolution of the luminosity function along redshift unnoticed during estimation.
 
For type classification based statistical functions, there is an additional complication 
that the $\xi_{Gg}$ for one class of galaxies might be very different with that for another class.
Furthermore it has been detected that color of galaxies, e.g. $g-r$, is strongly 
correlated even galaxies are at 
separation upto scales as large as $\sim 20h^{-1}$Mpc \cite{SkibbaSheth2009},
it is highly  
possible that samples of local galaxies with $z< \sim 0.007$ is biased more or
less in color.

\subsection{Two-point correlation function}

\begin{figure}
\epsfxsize=9cm
\epsffile{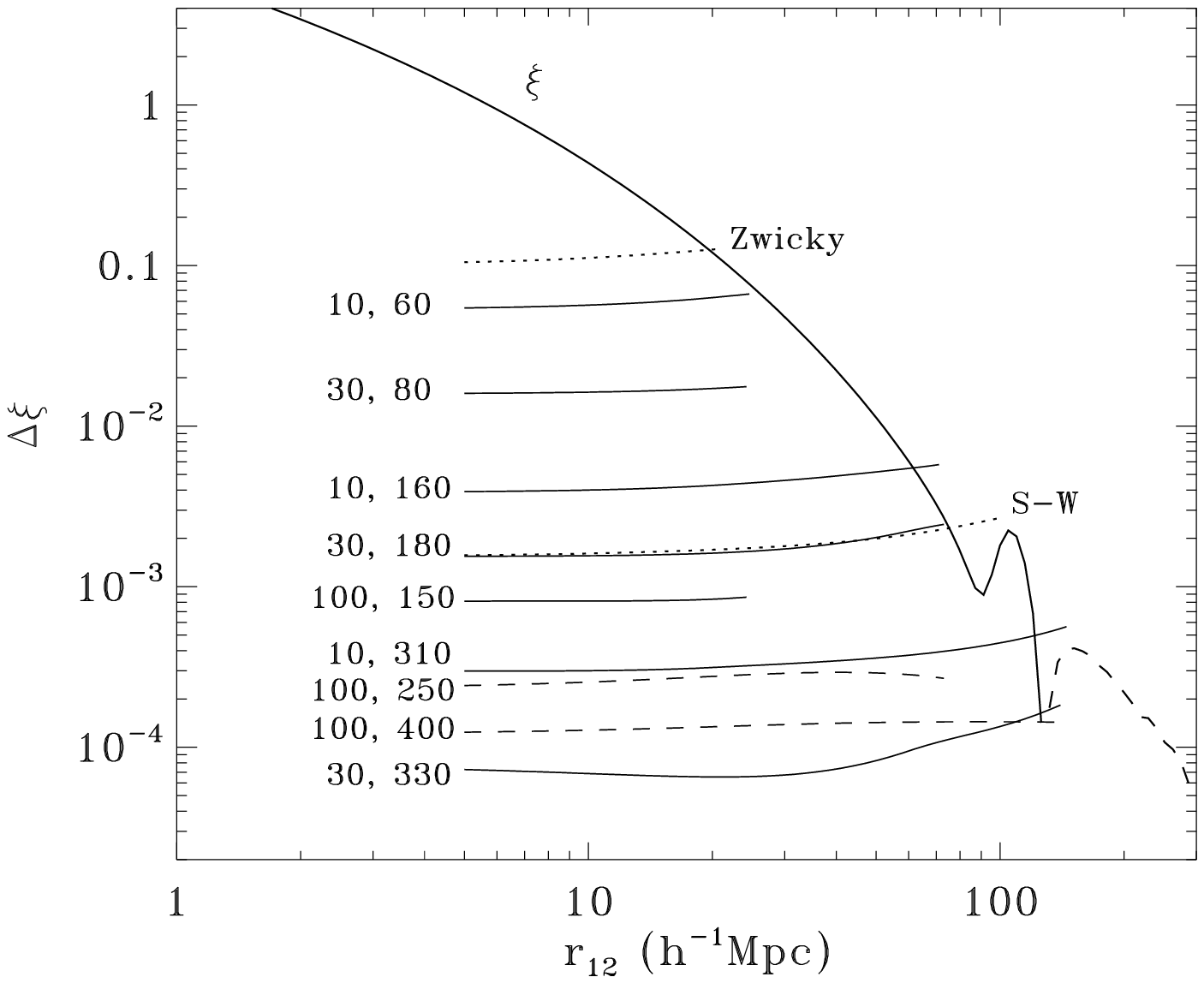}
\epsfxsize=9cm
\epsffile{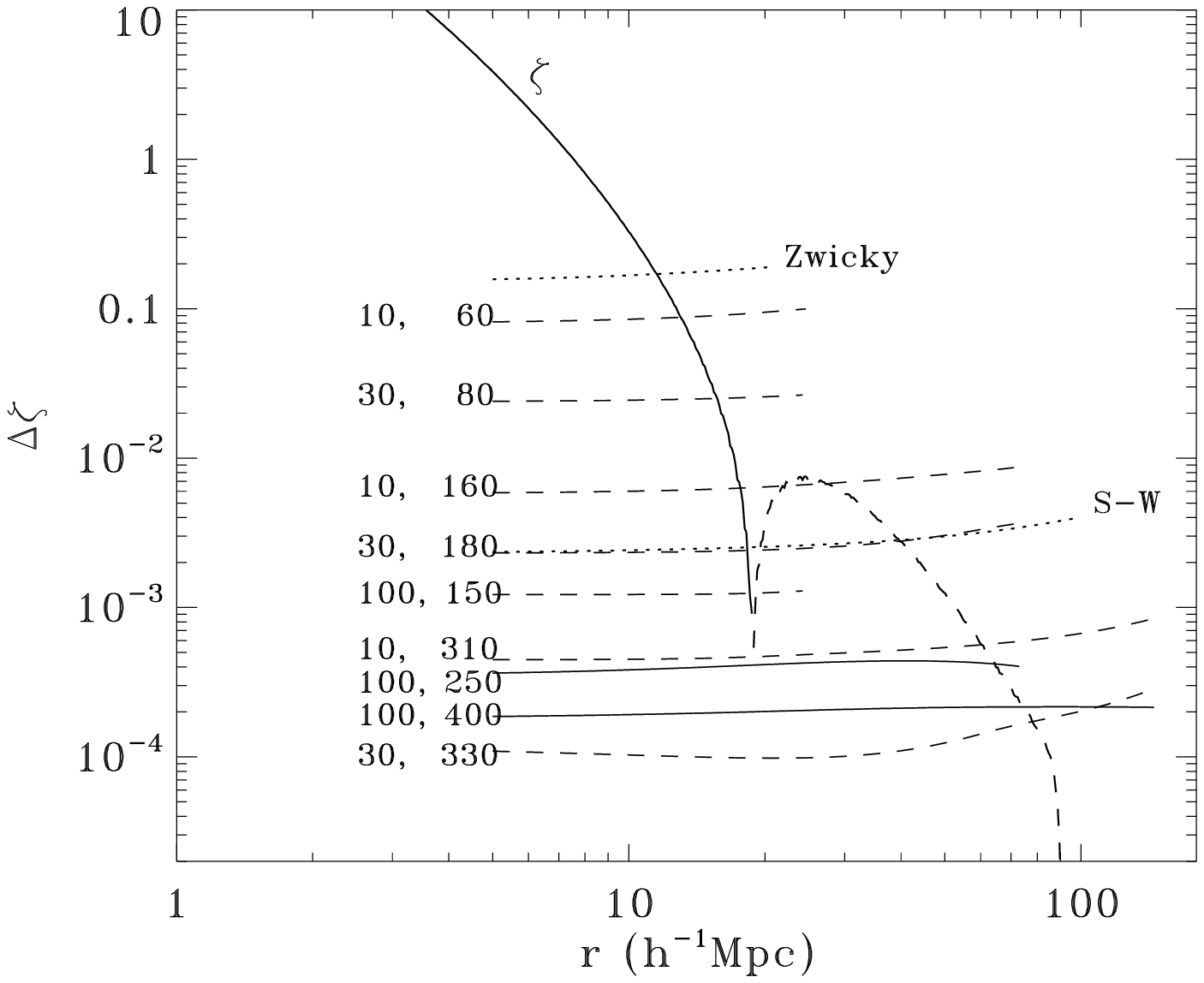}
\caption{Ergodicity biases in 2/3PCFs  of samples with different distance cuts. 
Those almost horizontal lines are $\Delta\xi$ and $\Delta\zeta$ 
given by Eq.~\ref{eq:dxi} and Eq.~\ref{eq:dzeta} respectively
provided that $b_{Gg}=1$, to the left ends of which are pairs of numbers labeling
distance cuts $(r_{\rm min}, r_{\rm max})$ of hypothetical samples. The
largest scale at which estimation of 2/3PCF is robust is chosen to be $(r_{\rm max}-r_{\rm min})/2$. 
Dashed lines refer to negative value. 
Left panel is of 2PCF while the right panel
displays the case of the 3PCF of equilateral 
configuration $\zeta(r_{12}=r_{23}=r_{31}=r)$. The top solid curve in the left plot
is the linear 2PCF at $z=0$ derived from the power spectrum provided by 
CMBFAST \cite{CMBFAST1996} with parameters 
$\Omega_m=0.27, \Omega_b=0.046, \Omega_\Lambda=0.73, \sigma_8=0.9, n=1$, and the $\zeta$ in the right
plot is the prediction of the Eulerian perturbation theory at tree-level \cite{BernardeauEtal2002}.
The dotted lines annotated with ``Zwicky'' approximates the
Zwicky catalogue which characteristic depth is $47.2h^{-1}$Mpc, while dotted lines coincident
with lines of (30, 180) but marked with ``S-W'' mimic the Shane-Wirtanen catalogue of characteristic depth
$209h^{-1}$Mpc \cite{GrothPeebles1977}, note that the two dotted lines in the right panel are actually
$-\Delta\zeta$.}
\label{fig:dxi}
\end{figure}

For an observer randomly placed in the Universe, the probability
of finding a pair of galaxies in two volume elements at positions ${\bf r}_1$ 
and ${\bf r}_2$ on ensemble average
is related to the two-point correlation function (2PCF) through
\begin{equation}
dP_2\propto \left[1+\xi_g(r_{12})\right]d{\bf r}_1 d{\bf r}_2\ ,
\label{eq:t2pcf}
\end{equation}
with $r_{12}=|{\bf r}_1-{\bf r}_2|$. 
It is this $\xi_g$ function that we aim at 
measuring, and shall be equal to the estimated $\hat{\xi}_g$ which is defined through
our observed possibility of finding pair of galaxies 
\begin{equation}
dP_2^{(O)}\propto [ 1+\hat{\xi}_g({\bf r}_1, {\bf r}_2) ] d {\bf r}_1 d{\bf r}_2\ .
\end{equation}

However, since we are the observer not randomly located but
in a galaxy as an object in the Universe, the observed probability of finding a pair 
of objects is in fact conditional to the object at origin point and shall be a 
three-points problem (see p. 173 of \cite{Peebles1980}),
\begin{equation}
dP_2^{(O)}\propto [ 1 + \xi_{Gg}(r_1)+ \xi_{Gg}(r_2)+
  \xi_g(r_{12})+\zeta_{Ggg}(r_1, r_2, r_{12}) ]  d{\bf r}_1 d{\bf r}_2
\label{eq:m2pcf}
\end{equation}
in which $\zeta_{Ggg}$ is the three-point cross correlation function.
The 2PCF we observed before averaging over ${\mathcal R}$ from galaxy sample 
evidently in principle is not the one in Eq.~\ref{eq:t2pcf} anymore but 
\begin{equation}
\hat{\xi}_g({\bf r}_1, {\bf r}_2)=\xi_g(r_{12})+\xi_{Gg}(r_1)+\xi_{Gg}(r_2)+
\zeta_{Ggg}(r_1, r_2, r_{12})\ ,
\label{eq:v2pcf}
\end{equation}
which can only be a good approximation to the targeted $\xi_g$ 
when $\xi_{Gg}(r_1)+\xi_{Gg}(r_2)+\zeta_{Ggg}(r_1, r_2, r_{12}) \ll \xi_g(r_{12})$. 
This, in together with
the fact that $\xi_{Gg}$ decreases with increasing distance and keeps
positive before zero-crossing, immediately
lets an amusing conclusion that galaxies close to us, on average, are clustered
more strongly than distant galaxies even if there are no evolutions resulted from 
gravitation force and galaxy bias function.

The measured 2PCF is actually averaged over the sample space ${\mathcal R}$
\begin{equation}
\hat{\xi}_g(r_{12})=\langle \hat{\xi}({\bf r}_1, {\bf r}_2)\rangle_{\mathcal R}=\frac{\int_{\mathcal R}\int_{\mathcal R} \hat{\xi}_g({\bf r}_1, {\bf r}_2)
\delta_D(|{\bf r}_1-{\bf r}_2|-r_{12})d {\bf r}_1 d{\bf r}_2}
{\int_{\mathcal R}\int_{\mathcal R}\delta_D(|{\bf r}_1-{\bf r}_2|-r_{12})d {\bf r}_1 d{\bf r}_2}\ ,
\label{eq:o2pcf}
\end{equation}
where $\delta_D$ is the Dirac delta function. 
It is difficult to provide exact figures about the errors by Eq.~\ref{eq:o2pcf}
before we acquire knowledge of the cross-correlation function
at two- and three-point level between the Galaxy and those observational selected 
sample galaxies. Since in practice galaxy samples' near-end distance limits are usually greater 
than $\sim 10 h^{-1}$Mpc, the regime
in consideration is fairly linear, we can comfortably assume that the bias 
of the Galaxy-galaxy cross-correlation
functions to the dark matter correlation functions is scale independent and linear, so that
\begin{equation}
\xi_{Gg}(r_{1,2})\approx b_{Gg}^2\xi(r_{1,2})
\ ,\ \ \xi_g(r_{12})=b_g^2\xi(r_{12}) \ ,\ 
\zeta_{Ggg}\approx b_{Gg}^2 b_g\zeta
\end{equation}
with $b_g$ being the bias of the sample galaxies to the dark matter and $b_{Gg}$ being the
bias of the Galaxy-galaxy correlation to the 2PCF of dark matter. 
In the weakly nonlinear regime
$\zeta\sim \xi^2$ while $\xi < 1$ and $b \sim 1-3$
(e.g. \cite{NorbergEtal2002a,TegmarkEtal2004a,WangEtal2007,SwansonEtal2008}),
the 3PCF term $\zeta_{Ggg} \ll \xi_{Gg}$ and goes to zero much faster than $\xi$ as 
scales increases, 
which thus can be ignored. Furthermore, there is a $1\leftrightarrow 2$ symmetry in
Eq. \ref{eq:o2pcf}, we then have 
\ba
\Delta \xi_g &=& \hat{\xi}_g(r_{12})-\xi_g(r_{12})\simeq \langle
\xi_{Gg}\rangle_{\mathcal R} \nonumber \\
&=&2\frac{\int_{\mathcal R}\int_{\mathcal R} \xi_{Gg}(r_1)
\delta_D(|{\bf r}_1-{\bf r}_2|-r_{12})d {\bf r}_1 d{\bf r}_2}
{\int_{\mathcal R}\int_{\mathcal R}\delta_D(|{\bf r}_1-{\bf r}_2|-r_{12})d
  {\bf r}_1 d{\bf r}_2 }\ .
\ea
We define a new function $\epsilon({\bf r}_1,r_{12})\equiv \int_{\mathcal R}
\delta_D(|{\bf r}_1-{\bf r}_2|-r_{12}) d{\bf r}_2/4\pi r_{12}^2$, which is the fraction
of surface area inside ${\mathcal R}$ of the sphere centered at ${\bf r}_1$ with 
radius $r_{12}$. If the survey
volume is sufficiently large that the boundary effect is negligible,
$\epsilon=1$. In general, $\epsilon$ depends on the survey geometry and can only be
evaluated numerically. However, under
the limit of full sky coverage, the analytical
expression of $\epsilon$ can be easily derived and $\epsilon({\bf
  r}_1,r_{12})=\epsilon(r_1,r_{12})$. We then use this approximation
\be
\Delta \xi_g\simeq 2\frac{\int_{r_{\rm min}}^{r_{\rm max}} \xi_{Gg}(r_1)
\epsilon(r_1,r_{12}) r_1^2dr_1}{\int_{r_{\rm min}}^{r_{\rm max}} \epsilon(r_1,r_{12})
  r_1^2dr_1}
\label{eq:dxi}
\ee 
to evaluate the ergodicity bias. 

Several numerical examples are demonstrated in Figure~\ref{fig:dxi}, 
the general trend of $\Delta\xi$ is that it trails off when $r_{\rm min}$ and
$r_{\rm max}-r_{\rm min}$ increases, cases in exception may occur when the
zero-crossing scale of 2PCF is between $r_{\rm min}$ and $r_{\rm max}$. (1) In the
limit that $r_{12}\ll (r_{\rm max}-r_{\rm min})/2$, $\epsilon(r_1,r_{12})\simeq 1$ for
most $r_1$ in the survey volume, thus $\Delta\xi$ is not sensitive to $r_{12}$
and to a good extent $\simeq 6b^2_{Gg}\int_{r_{\rm min}}^{r_{\rm max}}\xi r^2
dr/(r_{\rm max}^3-r_{\rm min}^3)$. As $\int_0^{\infty}\xi_{Gg}(r)r^2dr=0$ and
$\xi_{Gg}$ changes from positive to negative from small to large scales,
$\int_{r_{\rm min}}^{r_{\rm max}}\xi r^2 dr$ (and thus $\Delta \xi$) can 
deviate significantly from zero for 
some configurations of $[r_{\rm min},r_{\rm max}]$. However, the condition
$r_{12}\ll (r_{\rm max}-r_{\rm min})/2$ often means $r_{12}$ is small,
$\xi(r_{12})$ is large and thus $\Delta \xi\ll \xi(r_{12})$. (2) It looks that
when the characteristic redshifts are low and  $r_{12}\sim  (r_{\rm max}-r_{\rm min})/2$,
$\epsilon(r_1,r_{12})$ can considerably deviate from unity for many $r_1$ in
the survey volume and both $\Delta\xi$ and $\Delta \xi/\xi$ could
become significant,  but in this case the cosmic variance often overwhelms the
ergodicity bias.   (3) For deep surveys with $r_{\rm min}\gg r_c$,  the
ergodicity bias vanishes since $\Delta
\xi \sim 2\xi(r_{\rm min})\rightarrow 0$, where $r_c\simeq 120 h^{-1}$Mpc is
the zero point of the correlation function ($\xi(r_c)=0$). 
Thus it seems unlikely that the ergodicity bias can be significant in
practical means. 

\subsection{Three-point correlation function}
Similarly the observed probability of finding a triplet of galaxies is conditional 
to the Milk way and becomes a four-point problem
\ba
dP^{(O)}_3 &\propto & \left[ 1+ \xi_{Gg}(r_1)+\xi_{Gg}(r_2)+\xi_{Gg}(r_3)+
  \xi_g(r_{12})+\xi_g(r_{23})+\xi_g(r_{31})\right. \nonumber\\
  &+ & \xi_{Gg}(r_1)\xi_g(r_{23})+\xi_{Gg}(r_2)\xi_g(r_{31})+\xi_{Gg}(r_3)\xi_g(r_{12})\\
  &+ & \zeta_{Ggg}(r_1,r_{12},r_2)+\zeta_{Ggg}(r_1,r_{31},r_3)+\zeta_{Ggg}(r_2, r_{23}, r_3) + \zeta_g(r_{12},r_{23}, r_{31}) \nonumber\\
  &+ &\left. \eta_{Gggg}(r_1, r_2, r_3, r_{12}, r_{23}, r_{31}) \right] d{\bf r}_1 d{\bf
  r}_2 d{\bf r}_3 \nonumber
\label{eq:m3pcf}
\ea
in which $\eta_{Gggg}$ is the four-point cross-correlation function. 
The 3PCF we have is practically estimated via
\begin{equation}
\hat{\zeta}_g=X-\hat{\xi}_g(r_{12})-\hat{\xi}_g(r_{23})-\hat{\xi}_g(r_{31})-1\ ,
\end{equation}
where $X$ denotes the average of all those terms inside square brackets in Eq.~\ref{eq:m3pcf} over
sample space ${\mathcal R}$.
Substituting Eq.~\ref{eq:o2pcf} for $\hat{\xi}$ then yields
\begin{equation}
\hat{\zeta}_g=\zeta_g
+\langle \xi_{Gg} \rangle_{\mathcal R}  \left[\xi_g(r_{23})+\xi_g(r_{31})+  \xi_g(r_{12})-3\right]+ \langle \eta_{Gggg}\rangle_{\mathcal R} \ .
\label{eq:o3pcf}
\end{equation}
The ergodicity bias in the 3PCF is apparently much more difficult to analyze than the 2PCF due to its
complex configuration dependence. Nevertheless, if working on
large scales only where $\xi_g\ll 1$, those higher order terms can be neglected in Eq.~\ref{eq:o3pcf}, 
and dominant contribution just comes from the term $-3\langle \xi_{Gg}\rangle_{\mathcal R}$. As an order 
of magnitude estimation, the ergodicity bias in the 3PCF at
large scales is therefore roughly 
\begin{equation}
\Delta\zeta_g=\hat{\zeta}_g-\zeta_g \simeq -3\Delta \xi_g/2 \ . 
\label{eq:dzeta}
\end{equation}

It is known 3PCF approaches zero much faster than 2PCF when scale increases, the 
systematical bias identified here have much stronger effects to the third order 
statistical functions, which is obvious in the right panel of Fig.~\ref{fig:dxi}. Furthermore
as in most cases $\Delta\xi_g>0$ for local galaxy samples, the ergodicity bias in 3PCF
effectively behaves like a negative nonlinear galaxy bias parameter $b_2$ \cite{FryGaztanaga1993}, 
which imposes serious questions on the reliability of the nonlinear galaxy bias parameters
estimated through 3PCF of local galaxy samples and henceforth other related results.

\section{Discussion}
Here it is argued that by changing the point of view to that the observed
distribution of galaxies in the Universe is the distribution of neighbors to our Galaxy, 
statistics of the distribution are conceptually very different to what we used to think of,
though numerically the resulting ergodicity bias might
be small for most of practical applications especially when the galaxy sample is 
sufficiently far away from us and very deep.
Note that it has been assumed the correlation 
function between the Milk way and other galaxies follows the 
ensemble average $\xi_{Gg}$ and $\zeta_{Ggg}$, in reality the true
correlation strength could have large deviation to the mean since our Galaxy is
located on the outskirts of a large cluster, exact
numerical effects have to be explored carefully perhaps with the help of numerical
simulations.

Here we briefly discuss the impact of the ergodicity bias on precision
cosmology. (1) The baryonic acoustic oscillation (BAO) cosmology, which relies on 
the correlation measurement at
$r_{12}\simeq 100 h^{-1}$ Mpc. Mean redshifts of  galaxy samples constructed
for BAO detection in general are at  $z\sim 0.2$ or higher 
(e.g. \cite{PercivalEtal2010}) and thus $r_{\rm min}\gg r_c$.  We then
expect the ergodicity bias to  have little numerical
influence on the BAO detection. (2) The primordial
non-Gaussianity study through the galaxy power 
spectrum \cite{DalalEtal2008,SlosarEtal2008,Seljak2009} and  
bispectrum (e.g. \cite{JeongKomatsu2009}) at
scales even larger   than $100h^{-1}$Mpc . From
Fig.~\ref{fig:dxi}, we can conclude that the ergodicity bias certainly bias
their results. Precision measurements of the primordial non-Gaussianity require  
larger survey depth than we have
numerically evaluated, for which the induced bias is unlikely
significant, but may still be non-negligible. Especially, the method proposed
by \cite{Seljak2009} eliminates the cosmic variance in the power spectrum
measurement by taking the ratio of the power spectra of different
tracers.  Since taking ratio does not eliminate the additive ergodicity bias,
its relative impact is enhanced.  Robust evaluation of the
ergodicity bias in this case   
requires careful treatment of survey boundary, selection function and the
intrinsic evolution of galaxy number density and clustering. We leave this
detailed calculation elsewhere. 

In this short report only the impact on the spatial distribution of galaxies
is discussed as examples, there are possibly many other aspects of statistical 
analysis of galaxy samples 
in needs of similar conceptual adjustment. For instance the peculiar velocity
of galaxy we measured is actually the relative peculiar velocity of the 
galaxy to our Galaxy, and the peculiar velocities of galaxies are correlated with
the peculiar velocity of the Milky Way.

We must address that we are not challenging the Copernican Principle
and the Cosmological Principle here, but rather simply point out an observational 
effect. If there were observers who  
are randomly placed in the Universe, they will have the same conclusion as ours
about the sample provided by us. And the last thing we want to make clear is that
the correlation between the Galaxy and other galaxies
is not caused by our Galaxy, but is inherited from the intrinsic correlation
in the underlying dark matter distribution and the roughly synchronous evolution of these
galaxies.

\ack
The authors sincerely appreciate the many helpful advices of the anonymous referee. We also
would like to thank Jia-Sheng Huang, Lifan Wang and Zheng Zheng for enjoyable discussion
and Xianzhong Zheng for organizing the wonderful 4th Cosmology and Galaxy formation
summer workshop in which the work is concluded.
JP and PZ acknowledge the One-Hundred-Talent fellowships of CAS. JP and PZ are
supported by the Ministry of Science \& Technology of China
through 973 grant of No. 2007CB815401, 2007CB815402 and the NSFC through
grants of Nos. 10533030,  10633040, 10821302, 10873035.


\end{document}